\documentclass[preprint,showpacs,preprintnumbers,amsmath,amssymb]{revtex4}

\usepackage[dvips]{graphicx}
\usepackage{colordvi} 
\usepackage{dcolumn}
\usepackage{bm}
\usepackage{amsmath}

\renewcommand{\vec}[1]{\mbox{\boldmath $#1$}}
\begin{document}
\title{A Simulation Method to Resolve Hydrodynamic Interactions in Colloidal Dispersions}

\author{Yasuya Nakayama$^{1}$}
\email{nakayama@cheme.kyoto-u.ac.jp}
\author{Ryoichi Yamamoto$^{1,2}$}
\affiliation{
$^{1}$PRESTO, Japan Science and Technology Agency, 4-1-8 Honcho
Kawaguchi, Saitama 332-0012, Japan
}
\affiliation{
$^{2}$Department of Chemical Engineering, Kyoto University, 
Kyoto 615-8510, Japan
}

\date{\today}
\begin{abstract}
A new computational method is presented to resolve hydrodynamic
interactions acting on solid particles immersed in incompressible host
fluids.  In this method, boundaries between solid particles and host
fluids are replaced with a continuous interface by assuming a smoothed
profile.  This enabled us to calculate hydrodynamic interactions both
efficiently and accurately, without neglecting many-body interactions.
The validity of the method was tested by calculating the drag force
acting on a single cylindrical rod moving in an incompressible Newtonian
fluid.  This method was then applied in order to simulate sedimentation
process of colloidal dispersions.
\end{abstract}

\maketitle
\section{Introduction}

There are a number of useful systems consisting of small solid particles
dispersed in host fluids.  Among them, colloidal dispersions are most
common to our daily life and are of great importance, particularly in
the fields of engineering and
biology~\cite{russel89:_colloid_disper,larson99:_struc_rheol_compl_fluid}. Colloidal
dispersions have been reported to exhibit several unusual phenomena,
such as long-range correlations in sedimenting particles~\cite{segre01},
long-range anisotropic interactions in liquid crystal colloidal
dispersions~\cite{stark01:_physic}, transient gel formations during
phase separations of colloidal suspensions~\cite{tanaka99:_viscoel}, and
electro-rheological effects in particle suspensions of nonconductive
fluids~\cite{winslow49:_induc_fibrat_suspen}.

Since the dynamics of colloidal dispersions are very complicated, it is
extremely difficult to investigate their dynamic properties by means of
analytical methods alone.  Computational approaches are necessary in
order to elucidate the true mechanisms of dynamic phenomena in a variety
of situations.  Colloidal dispersions, however, have a typical
multi-scale problem.  The molecules comprising host fluids are much
smaller and move much faster than colloidal particles. From a
computational point of view, performing fully microscopic molecular
simulations for this kind of multi-scale system is extremely
inefficient.  An alternative, which is generally considered much better
than microscopic simulations, is to treat host fluids as coarse-grained
continuum media.

Several numerical methods have been developed in an effort to simulate
colloidal dispersions.  Two of the most well known methods are the
Stokesian dynamics~\cite{brady88:_stokes_dynam} and the
Eulerian--Lagrangian method.  The former is thought to be the most
efficient Method, capable of treating hydrodynamic interactions
properly. Furthermore, it can be implemented as $O(N_{p})$ scheme for
$N_{p}$ particles by utilizing the fast multipole
method~\cite{ichiki02:_improv_stokes_dynam}. However, it is extremely
difficult to deal with dense dispersions and dispersions consisting of
non-spherical particles by means of Stokesian dynamics due to the
complicated mathematical structures used in Stokesian dynamics.  On the
other hand, the Eulerian--Lagrangian method is a very natural and
sensible approach to simulate solid particles with arbitrary shapes.  A
number of kinds of tailor-made mesh, including unstructured mesh,
overset mesh, and boundary-fitted coordinates, have been applied to
specific problems, so that the shapes of the particles are properly
expressed in the discrete mesh-space.  Thus, in principle it is possible
to apply this method to dispersions consisting of many particles with
any shape.  However, a numerical inefficiency arises from the following:
i) re-constructions of the irregular mesh are necessary at every
simulation step according to the temporal particle position, and ii) the
Navier--Stokes equation must be solved with boundary conditions imposed
on the surfaces of all colloidal particles.  The computational demands
thus are enormous for systems involving many particles, even if the
shapes are all spherical.

Thus, our goal is to develop an efficient simulation method that can be
applied to particle dispersions in complex fluids.  Since host fluids
are considered incompressible in such systems, an efficient simulation
must address how to efficiently and accurately evaluate hydrodynamic
interactions.  As a first step towards this goal, we attempted to
develop a method to simulate colloidal dispersions in simple Newtonian
fluids.  The reliability of this method was tested by calculating the
drag force acting on a cylindrical object in a flow. Its performance was
subsequently demonstrated by simulating the sedimentation processes of
colloidal particles in a Newtonian fluid within a small Reynolds number
regime.

\section{Simulation method}
\begin{figure}[floatfix]
 \includegraphics[width=.5\hsize]{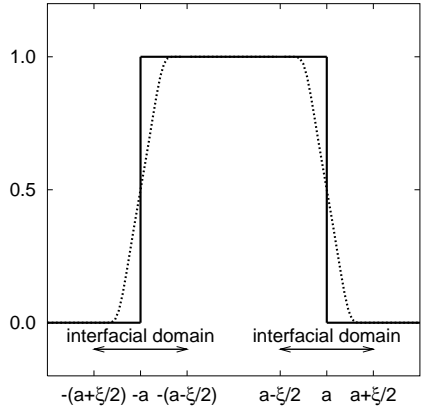} 
\caption{An example of the smoothed profile (dashed
line). The original rectangular profile is also shown
for comparison (solid line).}  
 \label{fig:profile}
\end{figure}

In order to overcome the problems arising at the solid-fluid interface
in the Eulerian--Lagrangian method, rather than the original
discontinuous rectangle profile (interfacial thickness, $\xi=0$)
schematically depicted in Fig.~\ref{fig:profile}, a smoothed profile was
introduced to the interface ($\xi>0$).  This simple modification greatly
benefits the performance of numerical computations, compared to the
original Eulerian--Lagrangian method for the following reasons.

i) Regular Cartesian coordinates can be used for many particle systems
with any particle shape, rather than boundary-fitted coordinates. The
solid-fluid interface has a finite volume ($\propto \pi a^{d-1}\xi$,
with $a$ and $d$ as the particle radius and system dimension) supported
by multiple grid points. Thus, the round particle shape can be treated
in fixed Cartesian coordinates without difficulty.  The simulation
scheme is thus free from the mesh re-construction problem that
significantly suppresses the computational efficiency of the
Eulerian--Lagrangian method. In addition, the simple Cartesian
coordinate enables use of the periodic boundary conditions as well as
the fast Fourier transformation (FFT).

ii) At the interfaces, the velocity component in the direction normal to
the interface of the host fluid must be equal to that of the particle.
This kinetic condition is imposed in the Navier--Stokes equation, as the
boundary value condition defined for the solid-fluid interface in
typical methods.  In our method, however, this condition is
automatically satisfied by an incompressibility condition on the entire
domain, which will be subsequently explained in detail.

iii) The computational demands for this method include sensitivity to
the number of grid points (volume of the total system), however it is
insensitive to the number of particles. Thus, our method is thought to
be suitable for simulating dense colloidal dispersions.

The non-zero interfacial thickness $\xi$ is the only approximation used
in the present method.  Thus, inter-particle hydrodynamic interactions
can be fully resolved within the approximation of the non-zero thickness
in the present method.

There have been two similar methods developed by different
authors~\cite{glowinski01:_fictit_domain_approac_direc_numer,hoefler00:_navier_stokes}.
The basic ideas of these methods is to use a fixed grid and represent
the particles not as boundary conditions to the fluid, but by a body
force or Lagrange multipliers in Navier--Stokes equation. The essential
difference in our approach to these two methods is to introduce a
explicit diffuse interface in a smoothed profile.  As long as a fixed
grid is used, the moving boundary is inevitably represented as diffuse
as grid spacing.  The introduction of a explicit smoothed profile makes
us to present a clear formulation of a numerical algorithm and is
advantagous when it is applied to complex
fluids~\cite{yamamoto01:_simul_partic_disper_nemat_liquid_cryst_solven,ryoichi04:_smoot_inter_method_simul_colloid_disper}

The lattice Boltzmann~(LB) method~\cite{succi01:_boltz} has attracted
much attention in recent years to simulate colloidal dispersions with
hydrodynamic interactions~\cite{ladd01:_lattic_boltz}.  The LB equation
was proved to offer a faithful discretization of Navier--Stokes
equation, and colloidal dispersions are simulated in the
Eulerian--Lagrangian manner. In practical viewpoints, the LB method is
forumlated on a fixed Cartesian lattice and is well adapted to parallel
computation. The LB approach, although the formulation is not intuitive
and its treatment of moving solid-fluid boundary is somewhat
complicated, has several similar merits to the present method.

The ``fluid particle dynamics''~(FPD) method was
proposed earlier and is similar to our method in
spirit~\cite{tanaka00:_simul_method_colloid_suspen_hydrod_inter}.  In
this method, although a similar smoothed profile was adopted, there are
several important differences between FPD and the present method.  The
most significant difference is that particles are modeled as a highly
viscous fluid with viscosity $\eta_{c}$, much greater than the fluid
viscosity $\eta_{s}$ in FPD. This enables the rigidity of the particles
to be sustained approximately by artificial diffusivity $\Delta
\eta\phi(\vec{x},t)$~($\Delta \eta \equiv \eta_{c}-\eta_{s} \gg
\eta_{s}$) within the particle domain. While this model is physically
correct, a practical problem remains in that a larger viscosity requires
smaller time increments.  In contrast, the present method treats
colloidal particles as undeformable solids, ({\itshape i.e.}, $\Delta
\eta \to \infty$) thus no additional constraint arises in the numerical
implementations.

\subsection{Basic working equations}

Colloidal dispersions are considered in a simple Newtonian liquid.  The
motion of the host fluid is governed by the Navier--Stokes equation with
the incompressibility condition,
\begin{eqnarray}
 \left(\partial_{t}+\vec{u}_{f}\cdot\nabla\right){\vec{u}_{f}} &=& \frac{1}{\rho}\nabla \cdot \vec{\sigma}_{f},
  \\
 \nabla \cdot \vec{u}_{f} &=& 0
\label{ns0}
\end{eqnarray}
where $\vec{u}_{f}$ is the fluid velocity, $\rho$ is the fluid
density. The stress tensor is represented by,
\begin{eqnarray}
 \vec{\sigma}_{f} &=& -p \vec{I} + \eta \left\{\nabla \vec{u}_{f}+ \left(\nabla \vec{u}_{f}\right)^{T}\right\}
\end{eqnarray}
where $p$ is the pressure, $\eta$ is the fluid viscosity.

The colloidal particles are assumed to be rigid and spherical, and their
positions $\vec{R}_{i}$ are tracked in a Lagrangian reference frame,
\begin{eqnarray}
 \dot{\vec{R}_{i}} &=& \vec{V_{i}},\label{eq:position}
\end{eqnarray}
with the translational momentum equation
\begin{eqnarray}
 M_{i}\dot{\vec{V}_{i}} &=& \vec{F}^{H}_{i} + \vec{F}^{PP}_{i} + \vec{F}^{g}_{i},
\end{eqnarray}
and the angler momentum equation
\begin{eqnarray}
 \vec{I}_{i}\cdot\dot{\vec{\Omega}}_{i} &=& \vec{N}^{H}_{i},\label{eq:torque}
\end{eqnarray} 
where $\vec{R}_{i}$, $\vec{V}_{i}$, $\vec{\Omega}_{i}$, $M_{i}$, and
$\vec{I}_{i}$ are the position, the translational velocity, the angular
velocity, the mass, and the inertia tensor of the $i$th particle,
respectively.  The hydrodynamic force $\vec{F}_{i}^{H}$ and torque
$\vec{N}_{i}^{H}$ acting on a particle can be obtained by integrating
the stress tensor over the surface as
\begin{eqnarray}
 \vec{F}^{H}_{i} &=& \int_{S_{i}}\vec{\sigma}_{f} \cdot d\vec{S} \label{eq:hydro_force0}
\\
 \vec{N}^{H}_{i} &=& \int_{S_{i}}\vec{r}\times\left(\vec{\sigma}_{f} \cdot d\vec{S}\right),
 \label{force0}
\end{eqnarray}
where $\vec{r}$ is the relative position vector from the center of
rotation to the colloid surface.  Furthermore, $\vec{F}^{PP}_{i}$ is the
force due to direct particle-particle interactions, and $\vec{F}^{g}_{i}
= M_{i}(1-\rho_{*}^{-1})\vec{g}$ is the buoyant force where
$\rho_{*}=\rho_{c}/\rho$ is the mass density ratio of the particles to
the host fluid, and $\vec{g}$ is the gravitational
acceleration. Relevant dimensionless parameters in the above equations
include the Reynolds number $Re=UL/\nu$, the Froude number $Fr=U/
\sqrt{gL}$, the mass density ratio $\rho_{*}$, and the
volume fraction $\alpha$.  Here $U$ and $L$ represent typical velocity
and length scales specific to the systems under consideration,
respectively.  The kinematic viscosity $\nu=\eta/\rho$ and the mass
density of colloidal particles $\rho_{c}$ are assumed to be constant.

In typical methods, the above set of equations should be solved using
proper boundary conditions defined at the solid-fluid interface.  In the
present method, however, the solid-fluid boundary condition is replaced
with a body force and an incompressibility condition on a total
velocity defined on the entire domain.

\subsection{Modified working equations}
  
Assuming a smoothed profile with a finite thickness $\xi$ to the
solid-fluid interface, we here derive the body force which
accurately takes the interactions between solids and fluids due to the
motions of colloids in an incompressible fluid into consideration.  The
present study considers a mono-disperse system consisting of $N$
spherical particles with radius $a$.  The positions of the particles
$\{\vec{R}_{1},\cdots,\vec{R}_{N}\}$ are first transformed to a
continuous field
\begin{eqnarray}
\phi(\vec{x},t) &\equiv& \sum_{i=1}^{N}\phi_{i}(\vec{x},t),\label{eq:phi}
\end{eqnarray}
using the $i$th particle's profile function $\phi_i(\vec{x})$ centered
at $\vec{R}_i$.  Several possible mathematical forms for
$\phi_i(\vec{x})$ exist, however, some typical functions are listed in
Appendix~\ref{appendix:smooth_profile}.

The continuum velocity field $\vec{u}_{p}$ is defined for the solid
particles using $\{\vec{V}_{1},\cdots,\vec{V}_{N}\}$,
$\{\vec{\Omega}_{1},\cdots,\vec{\Omega}_{N}\}$, and $\phi_{i}$ as
\begin{eqnarray}
 \phi \vec{u}_{p}(\vec{x},t) &\equiv& \sum_{i=1}^{N}
\left\{\vec{V}_{i}(t) + \vec{\Omega}_{i}(t)\times \left(\vec{x}-\vec{R}_{i}(t)\right)
\right\}
\phi_{i}(\vec{x},t).\label{eq:def_up}
\end{eqnarray}
The total (fluid+particle) velocity field is then given by
\begin{eqnarray}
 \vec{u}(\vec{x},t) &\equiv & (1-\phi) \vec{u}_{f} + \phi \vec{u}_{p} \nonumber
\\
&=& \vec{u}_{f} + \phi\left(\vec{u}_{p}-\vec{u}_{f}\right).\label{eq:total_u}
\end{eqnarray}
Since the particle velocity field $\vec{u}_{p}$ is constructed from the
rigid motions of particles, $\nabla\cdot\vec{u}_{p}=0$ is verified as
\begin{eqnarray}
 \nabla\cdot\vec{u}_{p} &=& 
  \nabla\cdot\sum_{i}\left\{
		      \vec{V}_{i}+\vec{\Omega}_{i}\times\left(\vec{x}-\vec{R}_{i}\right)
\frac{\phi_{i}}{\phi}
\right\}
\nonumber
\\
&=&
\sum_{i}\left\{
		      \vec{V}_{i}+\vec{\Omega}_{i}\times\left(\vec{x}-\vec{R}_{i}\right)\right\}
\cdot  \nabla
\frac{\phi_{i}}{\phi},
\\
\nabla
\frac{\phi_{i}}{\phi} &=& \frac{\left(\nabla\phi_{i}\right)\phi-\phi_{i}\left(\nabla\phi\right)}{\phi^{2}}
\nonumber
\\
&=& \frac{\left(\nabla\phi_{i}\right)\phi_{i}-\phi_{i}\left(\nabla\phi_{i}\right)}{\phi^{2}} = \vec{0}.
\end{eqnarray} 
Assuming the incompressibility of the fluid velocity
$\vec{u}_{f}$, the divergence of the total velocity is
\begin{eqnarray}
\nabla \cdot \vec{u} &=& \left(\nabla \phi\right)  \cdot(\vec{{u}}_{p}-\vec{{u}}_{f}).
\end{eqnarray}
The gradient of $\phi$ is proportional to the surface-normal vector and
have a support on the interfacial domains. Therefore, the
incompressibility condition on the total velocity $\nabla\cdot\vec{u}=0$
means the solid-fluid impermeability condition at the solid-fluid
interface.

We are to derive the evolution of the total velocity $\vec{u}$. To make
the points clearer, we first consider the problem assuming that the
motions of particles
$\left\{\vec{R}_{i}(t),\vec{V}_{i}(t),\vec{\Omega}_{i}(t)\right\}$ are
given. In Eq.~(\ref{eq:total_u}), only the fluid velocity $\vec{u}_{f}$
is to be solved. The evolution equation of the total velocity is
splitted as
\begin{eqnarray}
  \left(\partial_{t}+\vec{u}\cdot\nabla\right)\vec{u}&= & \frac{1}{\rho}\nabla \cdot \vec{\sigma},\label{eq:predict_u}
\\
\partial_{t}\vec{u} &=&  \phi \vec{f}_{p},\label{eq:fp}
\end{eqnarray}
where the stress tensor is
\begin{eqnarray}
 \vec{\sigma} &=& -p^{*} \vec{I} + \eta
  \left\{\nabla \vec{u}+ \left(\nabla \vec{u}\right)^{T}\right\}.\label{eq:stress}
\end{eqnarray}
By integrating Eqs~(\ref{eq:predict_u}) with (\ref{eq:stress}), the
total velocity is predicted as $\vec{u}=\vec{u}^{*}$. The pressure
$p^{*}$ in Eq.~(\ref{eq:stress}) is determined to fulfill the
incompressibility condition $\nabla\cdot\vec{u} = \nabla\cdot\vec{u}^{*}
= 0$.  Then, the body force $\phi\vec{f}_{p}$ is added to enforce
Eq.~(\ref{eq:total_u}) and solid-fluid inpermeability
condition. Therefore, the time-integrated body force $\phi\vec{f}_{p}$
is determined as
\begin{eqnarray}
 \int_{t}^{t+h}ds \phi\vec{f}_{p} &=& \phi\left(\vec{u}_{p}-\vec{u}^{*}\right) -\frac{1}{\rho}\nabla p_{p} h,\label{eq:int_fp}
\end{eqnarray}
where the pressure $p_{p}$ is determined to fulfill
$\nabla\cdot\vec{u}=\left(\nabla \phi\right)
\cdot(\vec{u}_{p}-\vec{u}^{*})=0$.  
By solving Eq.~(\ref{eq:fp}) with the body force, Eq.~(\ref{eq:int_fp}),
we finally make Eq.~(\ref{eq:total_u}) where the fluid velocity is
\begin{eqnarray}
 (1-\phi)\vec{u}_{f} &=& (1-\phi)\vec{u}^{*} -\frac{1}{\rho}\nabla p_{p}h.
\end{eqnarray}
We note that the non-slip
condition at the solid-fluid interface is fulfilled in this time
evolution of the total velocity.  Since the viscous
stress~(\ref{eq:stress}) acts on the entire domain including the
interfacial domain, the tangential velocity difference between
$\vec{u}_{f}$ and $\vec{u}_{p}$ is reduced. In other words, non-slip or
slip condition can be imposed in the definition of the stress
$\vec{\sigma}$ used in Eq.~(\ref{eq:predict_u}).
When compared to
FPD~\cite{tanaka00:_simul_method_colloid_suspen_hydrod_inter}, their
body force is $\phi\vec{f}_{p}=\Delta\eta\nabla\cdot \phi\left\{\nabla
\vec{u}+ \left(\nabla \vec{u}\right)^{T}\right\}$ with the artificial
diffusivity $\Delta \eta$. In the limit of $\Delta\eta \to \infty$, the
particle becomes rigid, however this limit cannot be achieved by the
numerical scheme used in FPD.  In contrast to FPD, the body force
$\phi\vec{f}_{p}$ guarantees the rigidity of the solid particle without
additional large artificial diffusivity.

We complete the time evolution by deriving the hydrodynamic force
$\vec{F}^{H}_{i}$ acting on the particles.
The hydrodynamic force is defined as the momentum flux between the fluid
and solid. Thus, 
the hydrodynamic force
is simply the counteraction from the fluid,
\begin{eqnarray}
 \vec{F}^{H}_{i} &=& -\int \rho\phi_{i}\vec{f}_{p}d\vec{x}.
\end{eqnarray}
In contrast to the force in Eq.~(\ref{eq:hydro_force0}) which is
expressed as the surface integral, the above force is given as the
volume integral. The volume integral is much advantageous in mesh-based
discretization compared to the surface integral since no generation of
body-fitted mesh is needed.

\subsection{Simulation procedure}
i) For a given particle configuration
$\{\vec{R}_{i}^{n}\}$, velocity $\{\vec{V}_{i}^{n}\}$,
and angular momentum $\{\vec{\Omega}_{i}^{n}\}$ (${i=1\ldots N}$), 
where the superscript $n$ denotes the time step and $h$ is the time 
increment, the fluid velocity at a time $t=n h$
is predicted as
\begin{eqnarray}
 \vec{u}^{*} &=& \vec{u}^{n-1} + \int_{t_{n-1}}^{t_{n-1}+h}ds \nabla \cdot \left(\frac{1}{\rho}\vec{\sigma} - \vec{u}\vec{u}\right)
,\label{eq:host_fluid_intermediate}
\end{eqnarray}
under the incompressibility condition $\nabla\cdot\vec{u}^{*}=0$ which
determines the intermediate pressure $p^{*}$.

ii) At step $n$, the total velocity $\vec{u}^{n}$ should be equal to
$\vec{u}_{p}$ within the particle domain and the surface-normal velocity
components of particles and fluid should be match in the interfacial
domain. Thus, it must be corrected by the body force $\vec{f}_{p}$
defined by
\begin{eqnarray}
\phi\vec{{f}}_{p}^{n} &=& \left\{\phi^{n}\left(\vec{u}_{p}^{n}-\vec{u}^{*}\right)\right\}/h-\frac{1}{\rho}\nabla p_{p}. \label{eq:def_fp}
\end{eqnarray}
The correcting pressure $p_{p}$ is determined to make a resultant total
velocity incompressible. This leads to the Poisson equation of $p_{p}$,
\begin{eqnarray}
 \nabla^{2}p_{p} &=& \rho\nabla\cdot\left\{\phi\left(\vec{u}_{p}^{n}-\vec{u}^{*}\right)\right\}/h.
\end{eqnarray}
Finally, we have
\begin{eqnarray}
 \vec{u}^{n} &=& \vec{u}^{*} + \phi \vec{f}_{p}^{n} h,\label{eq:host_fluid}
  \\
 p^{n} &=& p^{*} + p_{p}.
\end{eqnarray}

iii) The hydrodynamic force and torque acting on each colloidal particle
are now computed using the volume
integrals,
\begin{eqnarray}
 \vec{F}_{i}^{H} &=& -\int \rho\phi_{i}
\vec{f}_{p}^{n}d\vec{x}, \label{eq:hydro_force}\\
 \vec{N}_{i}^{H} &=& -\int \rho\left(\vec{x}-\vec{R}_{i}^{n}\right)\times\phi_{i}\vec{f}_{p}^{n}d\vec{x}.\label{eq:hydro_torque}
\end{eqnarray}
and the velocity, angular velocity, and position of each colloidal
particles at the current step $n$ are updated to $(n+1)$ as
\begin{eqnarray}
 \vec{V}_{i}^{n+1} &=& \vec{V}_{i}^{n} + \frac{1}{M_{i}}
\int_{t_{n}}^{t_{n}+h}ds
\left(
\vec{F}^{H}_{i} + \vec{F}^{PP}_{i} + \vec{F}^{g}_{i}
\right),\label{eq:vi}
\\
 \vec{\Omega}_{i}^{n+1} &=& \vec{\Omega}_{i}^{n} + \vec{I}_{i}^{-1}\cdot
\int_{t_{n}}^{t_{n}+h}ds
\vec{N}^{H}_{i},\label{eq:Omegai}
\\
 \vec{R}_{i}^{n+1} &=& 
 \vec{R}_{i}^{n} +
\int_{t_{n}}^{t_{n}+h}ds \vec{V}_{i}.\label{eq:ri}
\end{eqnarray}
Since the same $\phi\vec{f}_{p}$ is used for both the host fluid
(\ref{eq:host_fluid}) and the colloidal particles
(\ref{eq:hydro_force}), (\ref{eq:hydro_torque}) through the interface,
no excess or shorts for solid-fluid interactions exist. The same type of
solid-fluid interaction has been previously proposed in
Ref.~\cite{kajishima01:_turbul_struc_partic_laden_flow}, though the
treatment used in the present method is more general.  It is important
to note that the timing of the updates for fluid
(\ref{eq:host_fluid_intermediate}) and particles
(\ref{eq:vi})-(\ref{eq:ri}) has been shifted; the particles always go
one step ahead of the fluid. This shift is primarily due to a technical
issue related to the solid-fluid boundary condition. In general, the
boundary value conditions, which is replaced with the force due to
solid-fluid interactions in the present method, is necessary to update
the fluid. Otherwise, in order to update both the fluid and the
particles simultaneously, the implicit scheme for particles must be
used. 
Consider the problem to integrate the $(n-1)$th step variables,
$\{\vec{R}_{i}^{n-1},\vec{V}_{i}^{n-1}, \vec{\Omega}_{i}^{n-1}\}$ with
$\vec{u}^{n-1}$, to $n$th step. The particle velocity of next step,
$\vec{u}^{n}_{p}$, in Eq.~(\ref{eq:def_fp}) is needed to update the
total velocity $\vec{u}^{n-1}$ to $\vec{u}^{n}$ before
$\{\vec{R}_{i}^{n-1},\vec{V}_{i}^{n-1}, \vec{\Omega}_{i}^{n-1}\}$ is
updated. This situation requires the implicit treatment.
The use of the implicit scheme complicates the algorithm and reduces the
efficiency.  The timing-shift is therefore necessary to realize the full
explicit scheme described above. 
The initial condition,
$\{\vec{R}_{i}^{n},\vec{V}_{i}^{n}, \vec{\Omega}_{i}^{n}\}$ with
$\vec{u}^{n-1}$, can be generically constructed. From the total velocity
$\vec{u}^{n-1}$ satisfying the initial boundary condition given by
$\{\vec{R}_{i}^{n-1},\vec{V}_{i}^{n-1},\vec{\Omega}_{i}^{n-1}\}$, the
hydrodynamic force and torque is computed as surface integrals and the
particle trajectory is integrated from $(n-1)$th to $n$th step.
Together with the construction of the initial condition, the
timing-shift algorithm is realized without loss of generality.

In the above formulation, although the solid-fluid interaction
force~(\ref{eq:def_fp}) was computed on the basis of the Euler scheme
for simplicity of presentation, implementations of higher order schemes
are straightforward for Eqs.~(\ref{eq:vi})-(\ref{eq:ri}).  Furthermore,
in order to update the host fluid in
Eq.~(\ref{eq:host_fluid_intermediate}), no restrictions exist for the
time discretization and any conventional scheme can be used for
incompressible fluids.

\section{Numerical results}
The present method has been applied to two specific problems:
The calculation of the drag force acting on an infinitely long cylindrical rod 
moving in a Newtonian fluid in order to check the validity.
The method was also applied to simulations of many sedimenting particles in a two dimensional fluid in order 
to demonstrate the performance. 
In the present simulations, the Navier--Stokes equation was discretized with 
a de-aliased Fourier spectral scheme in space and a second order 
Runge--Kutta scheme (the Heun scheme) in time.  
For the colloidal particles, the velocity and the angular velocity were
integrated with the Heun scheme, and the position was integrated with 
the Crank--Nicolson scheme.
The external boundary condition on the edge of the systems was imposed in
the same manner as the fluid-solid boundary condition on the particle surface. 
The simulation code is remarkably simple due to such unified treatment for all boundary conditions.

\subsection{Drag force on a cylindrical rod}

\begin{figure}[floatfix]
\includegraphics[width=.5\hsize]{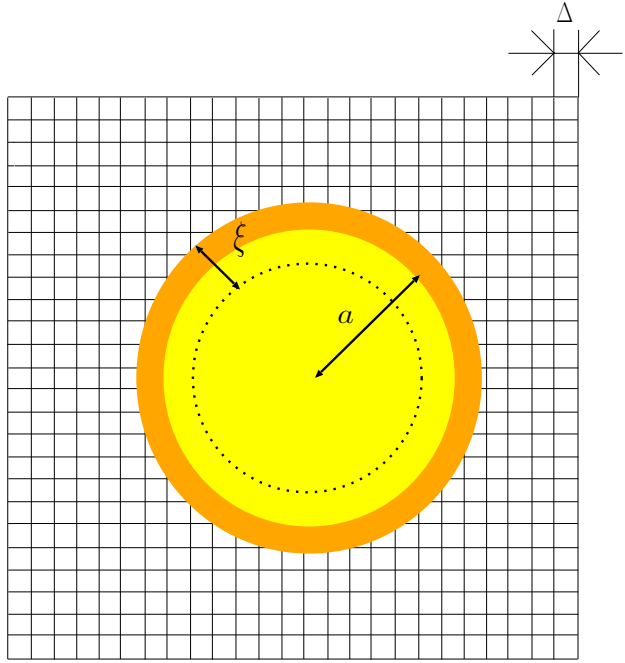}
 \caption{Schematic representation of the cross section of geometry around
 the rod. $\Delta$ is the lattice spacing, $a$ is the rod radius, and
 $\xi$ is the interfacial thickness.  The rod surface now has a finite
 volume $\sim 2\pi a\xi$ supported by several grid points on
 the fixed Cartesian coordinate.}
  \label{fig:schematic}
\end{figure}

The drag force acting on an
infinitely long cylindrical rod with radius $a$ was computed by
solving the Navier--Stokes equation around the rod in order to check the accuracy of the present method.
Figure~\ref{fig:schematic} shows a cross
section of the geometry around the rod with finite thickness $\xi$
at the interface.

\begin{figure}[floatfix]
\includegraphics[width=.8\hsize]{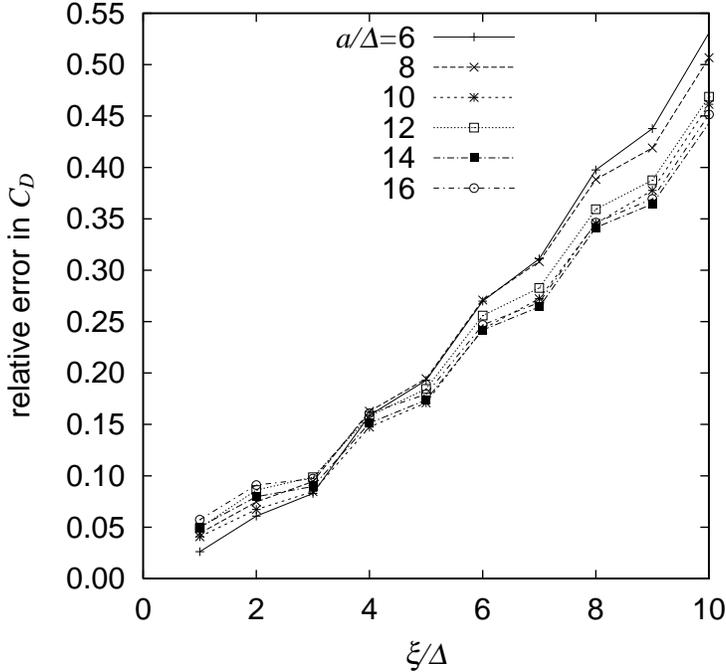}
\caption{The relative error in the drag coefficient $C_{D}$ 
as a function of the interfacial thickness $\xi/\Delta$. 
}
\label{fig:xidx_rel} 
\end{figure}

First, the effects of the finite thickness on the drag force are
examined in the square domain of $L^{2}$.  An uniform stream $U$ in
$x$-direction was assigned to the edge of the domain as the boundary
condition. Here the Reynolds number was defined by $Re=2aU/\nu$.  The
drag coefficient was calculated as $C_{D} = F_{D}/\rho U^{2} a$, where
the drag force $F_{D}$ was computed from Eq.~(\ref{eq:hydro_force}) for
various values of $\Delta$, $a$, $L$, and $U$.
Figure~\ref{fig:xidx_rel} shows the relative error
$\left\{C_{D}(Re,\xi/\Delta)-C_{D}(Re,\xi/\Delta=0)\right\}/C_{D}(Re,\xi/\Delta=0)$
as a function of the interfacial thickness $\xi/\Delta$, where
$C_{D}(Re,\xi/\Delta =0)$ was estimated by extrapolating the measured
curve of $C_{D}(Re,\xi/\Delta)$ to $\xi/\Delta \to 0$. The relative
error in $C_{D}$ was observed to increase with increasing $\xi/\Delta$,
however, it tended to converge within $5\%$ for several values of
$a/\Delta $ for $\xi/\Delta = 1$ and $0<Re<20$. Thus, $\xi/\Delta =1$
was set for further simulations.

\begin{figure}[floatfix]
 \includegraphics[width=.8\hsize]{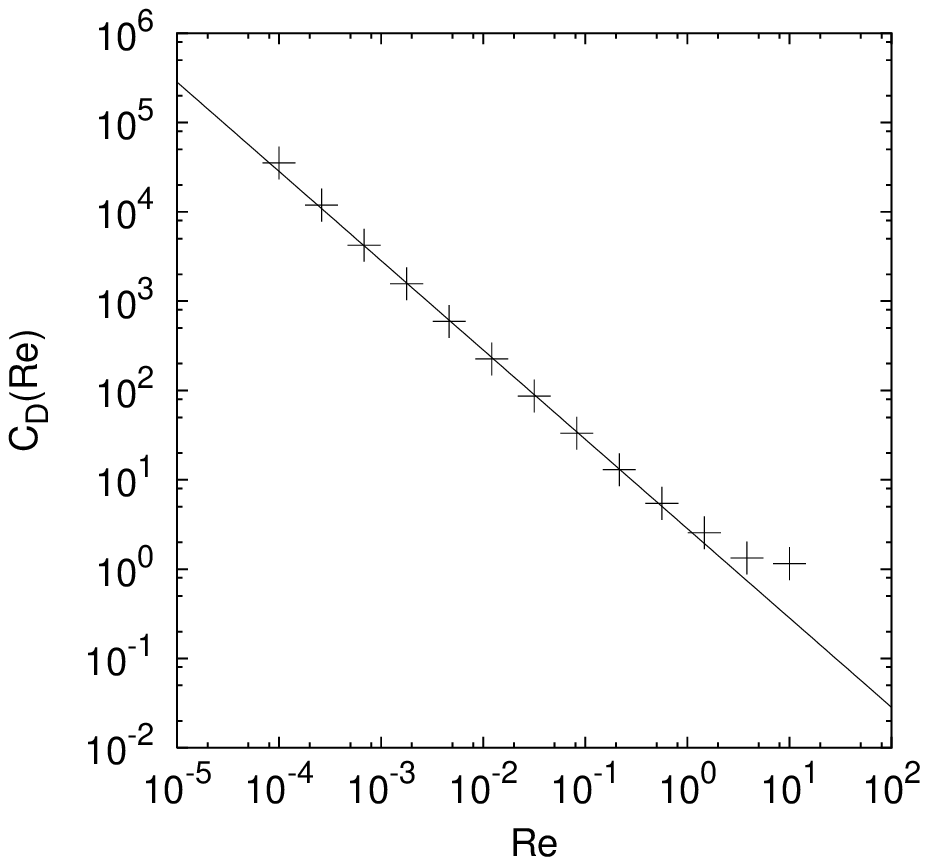}
 \caption{Comparisons of the drag coefficient $C_{D}$~(plus) from our
 method with the theoretical curve of the Stokes law~(solid line). }
\label{fig:CDR}
\end{figure}

Next, the drag coefficient $C_{D}$ was calculated.  The rod was fixed at
the origin in the circular domain with radius $L$.  The velocity at the
external boundary $r=L$ was set to $\vec{u}(r=L,\theta) = U/(1-2\log
(a/L))
\left[\left\{1-\left(a/L\right)^{2}-2\log\left(a/L\right)\right\}\vec{e}_{x}
-2\left\{1-\left(a/L\right)^{2}\right\}\cos\theta \vec{e}_r\right]$
where $\vec{e}_{x}$ and $\vec{e}_{r}$ are the unit vectors in $x$- and
$r$- directions, respectively, and $\tan\theta = y/x$. An analytical
solution for the Stokes equation is known for this boundary condition,
and the drag force is given by $F_{D}=8\pi\eta U/(1-2\log(a/L))$.  The
computed $C_{D}$ using the present method as a function of $Re$ is shown
in Fig.~\ref{fig:CDR}, and is in good agreement with the theoretical
Stokes law in $Re \leq 1$ within $5\%$.

The accuracy of the present method using the finite interfacial
thickness $\xi/\Delta = 1$ was determined to be acceptable for
simulating colloidal dispersions for $Re\le1$ based on the numerical
results above.

\subsection{Sedimentation}

\begin{figure}[floatfix]
 \includegraphics[width=.8\hsize]{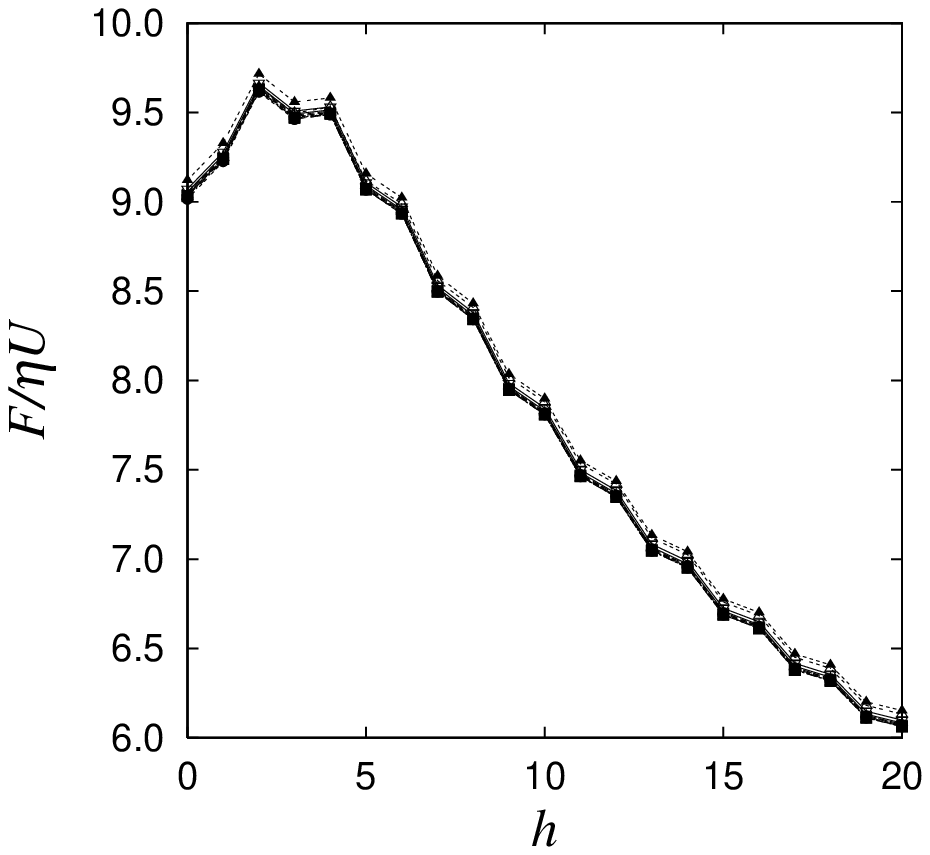}
 \caption{
The normalized lubrication force acts on two approaching infinitely long 
cylindrical rods as a function of the nearest distance between the two 
surfaces. Different symbols denote different approaching velocities $U$,
ranging from $2.5\times 10^{-6}$ to $9.6\times 10^{-2}$ by $U = 2.5\times 10^{-6}\times 2.61^{n},~ n=0\ldots 11$, 
which almost collapsed.  The observed scaling behavior $F=\eta U f(h)$
with scaling function $f(\dots)$ is characteristic of Stokes flow, due
to Reynolds numbers $2aU/\nu<1$.}  
\label{fig:lub}
\end{figure}

The performance of the present method was examined by simulating
sedimentation processes of mono-disperse particles in a two dimensional
Newtonian fluid in a rectangular box surrounded by non-slip
walls with $\rho_{*}=1.1$ and $\alpha=0.143$ The dimensionless
parameters were taken to be $Re=0.0916$, $Fr=0.0512$, where the settling
velocity and the diameter of particle were taken as the characteristic
velocity and length.  Other computational parameters were chosen as
$\Delta=1$, $\xi/\Delta=1$, $a/\Delta=10$, $L_{x}/\Delta = 512$, and
$L_{y}/\Delta = 1024$, where $y$-axis is in the direction of gravity. In order to prevent the particles from overlapping within the
core radius $\simeq a$, the force was added
$
\vec{F}_{i}^{PP} =
-\partial E_{PP}/\partial \vec{R}_{i}
$
due to direct particle-particle interaction
using the repulsive part of the Lennard-Jones potential 
$
E_{PP}=0.4\sum_{i=1}^{N-1}\sum_{j=i+1}^{N}
\left[\left(2a/R_{ij}\right)^{12} - \left(2a/R_{ij}\right)^{6}
\right]\theta(2^{7/6}a-R_{ij}),
$
where $\theta(\cdots )$ is the step function and $R_{ij}\equiv
\left|\vec{R}_{i}-\vec{R}_{j}\right|$.  The direct interaction
$\vec{F}_{i}^{PP}$ is not very important when the particles are moving
around because the particles never overlap due to the lubrication effect,
even without $\vec{F}_{i}^{PP}$.  
Figure~\ref{fig:lub} shows the lubrication force acting on two
approaching rods computed using the present method. The
lubrication force is always repulsive in this case, and thus prevents the
rods from approaching each other. The strength of the repulsion increases with increasing velocity $U$.
On the other hand, when the particles are stacked on the bottom wall during 
the later stage of sedimentation, $\vec{F}_{i}^{PP}$ is required to
sustain the stacking against gravity.  In fact, the repulsion vanishes
for immobile pairs of rods.

\begin{figure}[floatfix]
\vspace*{3pt}
 \includegraphics[width=.3\hsize,angle=0]{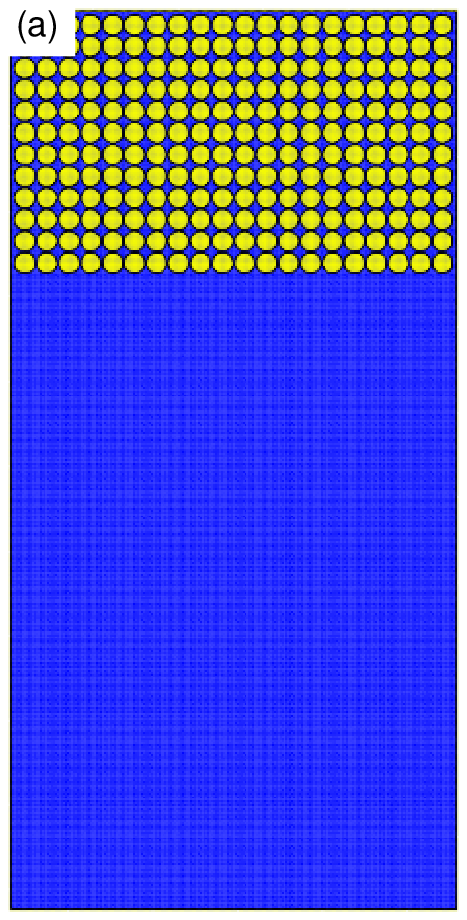}
 \hspace*{.2\hsize}
 \includegraphics[width=.3\hsize,angle=0]{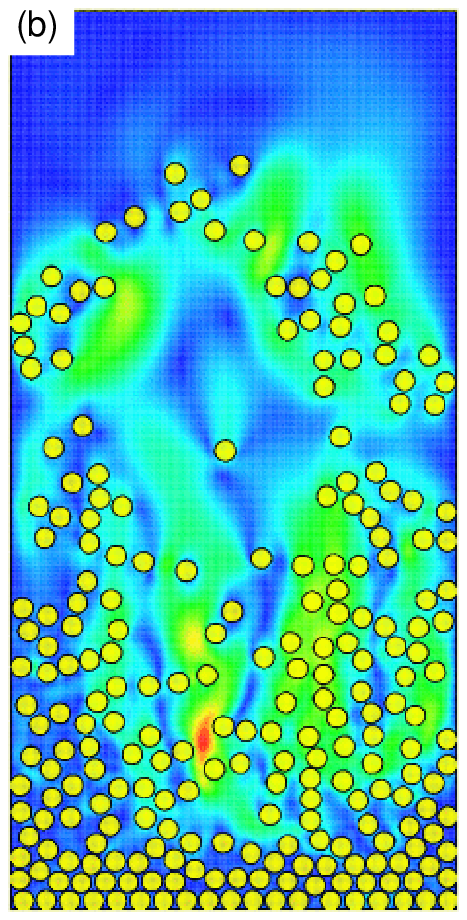}
 \caption{Snapshots of 240 colloidal disks sedimenting in a
 two-dimensional Newtonian fluid obtained using the present method. The
 magnitude of the host fluid velocity is indicated in color; change of
 color from blue to red corresponds to change of the fluid velocity from
 small to large.}

\label{fig:sedi}
\end{figure}

At the initial configuration, all the particles were placed near the
upper wall and both fluid and particle velocities were set to zero, as
depicted in Fig.~\ref{fig:sedi}(a). A typical snapshot during
sedimentation is shown in Fig.~\ref{fig:sedi}(b). Regions with swirled
particles were observed, in which the particle velocities were highly
correlated as a result of long-range inter-particle hydrodynamic
interactions.  A simulation with periodic boundary conditions in the
horizontal ($x$-)direction was also conducted.  In this simulation,
swirls were still developed, however they were smaller than those
observed with non-slip walls. The effect of confinement in the non-slip
walls therefore enhances the velocity correlation.  The computational
demand required for the present simulation is less than one day of
processing on a normal PC.

\section{Concluding Remarks}

A new computational method has been developed to simulate particle
dispersion in fluids. Utilizing a smoothed profile for solid-fluid
boundaries, hydrodynamic interactions in many particle dispersions can
be taken fully into account, both accurately and efficiently. In
principle, the present method can be easily applied to systems
consisting of many particles with any shape.  The reliability of the
method was examined by calculating the drag force acting on a
cylindrical object in a flow.  The performance of the method was
demonstrated to be satisfactory by simulating sedimentations of
particles in a Newtonian fluid.

Another primary benefit of using the smoothed profile arose when the
method was extended to colloidal dispersions in complex fluids with an
internal degree of freedom, such as the molecular orientation or ionic
density.  In complex fluids, inter-particle interactions can be mediated
by the internal degree of freedom of the fluid.  In such cases, the
fluid-particle interactions at the colloid surface could be more
efficiently handled by utilizing a smoothed profile. Previous studies on
particle dispersions in liquid crystal solvents demonstrate a striking
example of this
efficiency~\cite{yamamoto01:_simul_partic_disper_nemat_liquid_cryst_solven,ryoichi04:_smoot_inter_method_simul_colloid_disper}.
Although the hydrodynamic effects were neglected in these simulations,
extensions to implement the hydrodynamic effects by incorporating the
present method are currently underway.

\appendix
\section{Selection of smoothed profiles}
\label{appendix:smooth_profile}

The specific form of the smoothed profile should be selected according
to the convenience of the physical modeling of systems under
consideration.  In the present study, an infinitely differentiable
function with compact support was used. We adopted $\phi$ defined as
\begin{eqnarray}
 \phi_{i}(\vec{x}) &=& g(\left|\vec{x}-\vec{R}_{i}\right|),\\
 g(x) &=& \frac{h((a+\xi/2)-x)}{h((a+\xi/2)-x)+h(x-(a-\xi/2))},
  \\
 h(x) &=& \left\{
	   \begin{array}{cc}
	    \exp\left(-\Delta^{2}/x^{2}\right) & x \geq 0,
	     \\
	    0& x < 0.
	   \end{array}
\right.
\end{eqnarray}
where $\vec{R}_{i}$, $a$, $\xi$, and $\Delta$ were the position of the
particle, the radius of the particle, the interfacial thickness, and
lattice spacing, respectively.  This choice is shown in
Fig.~\ref{fig:schematic}. While this $\phi$ may appear somewhat complicated compared to other more simple choices, this $\phi$ has the following benefits: i) three domains;
solid, fluid, and interface are explicitly separated, namely, $\phi =
1$ is the solid domain ($\left|\vec{x}-\vec{R}_{i}\right| < a-\xi/2$), $\phi=0$ is the fluid domain ($a+\xi/2 < \left|\vec{x}-\vec{R}_{i}\right|$), and $0<\phi<1$ is the
interfacial domain ($a-\xi/2 < \left|\vec{x}-\vec{R}_{i}\right| < a+\xi/2$), ii) high order derivatives of $\phi_{i}$ with
respect to $\vec{x}$ can be analytically calculated, and iii) due to its
support-compactness, the integrals in Eqs.~(\ref{eq:hydro_force}) and
(\ref{eq:hydro_torque}) remain local, which contributes greatly to the
efficiency of the computation.

The second possible choice is
\begin{eqnarray}
 \phi_{i}(\vec{x}) &=& \frac{1}{2}\left(\tanh\frac{a-\left|\vec{x}-\vec{R}_{i}\right|}{\xi}+1\right).
\end{eqnarray}
This choice was used in
Refs~\cite{tanaka00:_simul_method_colloid_suspen_hydrod_inter,yamamoto01:_simul_partic_disper_nemat_liquid_cryst_solven,ryoichi04:_smoot_inter_method_simul_colloid_disper}.
This $\phi$ is also infinitely differentiable as well as analytically
easy to handle. However, the support is not compact and the separation
of the three domains is ambiguous.  Furthermore, the ambiguity of domain
separation tends to be more enhanced for higher order derivatives.  For
practical implementation, due to exponential decay of the hyperbolic
function, a proper cutoff radius is adopted for the calculation of the
integrals in Eqs.~(\ref{eq:hydro_force}) and (\ref{eq:hydro_torque}).

The third possible choice is given by
\begin{eqnarray}
 \phi_{i}(\vec{x}) &=& s(a-\left|\vec{x}-\vec{R}_{i}\right|),\\
 s(x) &=& \left\{
	   \begin{array}{cc}
	    0 & x < -\xi/2,
	     \\
	    \frac{1}{2}\left(\sin\frac{\pi x}{\xi}+1\right) & \left|x\right|<\xi/2,
	     \\
	    1 & x > \xi/2.
	   \end{array}
\right.
\end{eqnarray}
which has the property of exact separation of the three domains,
however, the second derivative of $\phi$ is discontinuous at the
fluid-interface boundary.  Therefore, it is not recommended for
computational models requiring derivatives higher than the second
order of $\phi$.

The detailed choice of $\phi$ does not affect the results of the present
simulations because only the first order derivative of $\phi$ is
required in the present case. However, care must taken if higher order
derivatives are required.

\end{document}